\documentclass[aps,prl,twocolumn,superscriptaddress,showpacs]{revtex4}
\usepackage{amsfonts}
\usepackage{graphicx}
\newcommand{\bea}{\begin{eqnarray}}
\newcommand{\eea}{\end{eqnarray}}
\newcommand{\figwidth}{0.7\columnwidth}
\begin{document}
\title{Fermions in optical lattices near  a Feshbach resonance:\\ from 
band insulator to Mott insulator}
\author{A.~F. Ho}
\affiliation{School of Physics and Astronomy,  The University of Birmingham,
Edgbaston, Birmingham B15 2TT, UK \\
Donostia International Physics Center (DIPC) Manuel de Lardizabal 4, 20018-Donostia, Spain \\
Department of Physics, Blackett Laboratory, Imperial College,
 Prince Consort Road, London SW7 2BW, UK.}
%
\begin{abstract}
We study a model of an equal mixture of two species of fermions in a deep optical lattice at a filling of two fermions per site. At weak inter-species interaction, the system is a band insulator.
When the inter-species interaction is tuned via a Feshbach resonance to be larger than an energy related to the
energy separation of the first and second Bloch band, atoms populate equally the two Bloch bands. With weak tunneling between sites of the optical lattice, the system becomes a Mott insulator with the low energy effective Hamiltonian of a spin-1
Heisenberg antiferromagnet, because of a Hund's rule like coupling between the
two bands. We discuss experimental signatures of these two types of insulators.
\end{abstract}
\date{\today}
\maketitle

The controllability and the capability to continuously tune parameters  has provided
unprecedented opportunities to study strong correlation physics in trapped 
ultra cold atoms systems.
Thus, Greiner {\it et al.}\cite{G02} observed the quantum phase transition from a superfluid to a Mott insulator for bosons trapped in an optical lattice, by suppressing tunneling between sites of the lattice.
Recently, there has been much experimental progress in studying fermionic superfluidity in the BCS to BEC  crossover regime \cite{BCSBEC} whereby the scattering length between two species of fermions is continuously tuned through a Feshbach resonance.\cite{BCSBECtheory}

In an optical lattice, the hopping matrix element $t$ and the on-site interaction 
strength $U$ both depend essentially on the amplitude $V_0$ of the laser beams that define the optical lattice.\cite{Jaksch98}
But by adding an appropriate magnetic field to sweep through a Feshbach resonance, a larger  range of $U$ vs. $t$ can be accessed experimentally. Very recently,
K{\"o}hl {\it et.al.}\cite{K04} have exploited this to study  two hyperfine states of 
fermionic $^{40}$K  (about 50:50 mixture) loaded into a three dimensional (3D) 
optical lattice. Initially, far from the Feshbach resonance ({\it i.e.} no interaction), the lattice is loaded with
two fermions per site (one for each spin state) into the lowest Bloch band, with weak tunneling between
sites. Hence the system is a band insulator. Now turning up the magnetic field towards the inter-species Feshbach resonance causes $U$ to increase to bigger than the interband energy, hence they measure
atoms being kicked into upper Bloch bands. The key question\cite{Jason} is this: what happens to the band insulator as the Feshbach resonance is approached?

For atoms with Feshbach resonance enhanced interactions, Diener and Ho\cite{Jason} have shown in the single site problem, and  Katzgraber {\it et.al.}\cite{Troyer05} in the full lattice problem,  that there can be a window of magnetic field where essentially 
two Bloch bands  are occupied: the harmonic oscillator ground state [000], together with
the [001] band for an anisotropic lattice (as in the experiment of \cite{K04}). At or very near the Feshbach resonance itself, more bands may be occupied.

\begin{figure}
 \begin{center}
 \includegraphics[width=\columnwidth]{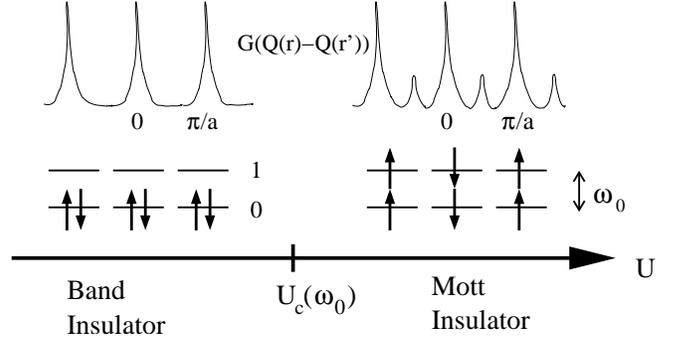}
 \caption{Schematic phase diagram. The middle row shows the dominant spatial structure of the two species ("spin") of fermions in orbital 0 and 1; the top row shows the schematic correlation function $G(r,r')$ that may be measured in noise correlation experiments: the Mott insulator has extra peaks due to doubling of real space unit cell.}\label{fig:phase}
 \end{center}
\end{figure}

Thus, motivated by the experiment of \cite{K04}, 
we study in this paper a simple model of spin-1/2 fermions in an optical lattice at a filling
of two fermions per site, with equal mix of the two spin species, and
where two Bloch bands are active, as a function of $U$:
\begin{eqnarray} \label{H}
H &=& \sum_{\alpha,\beta} \sum_{\sigma, i} -t_{\alpha \beta} \;
c^{\dagger}_{i+1 \sigma \alpha} c_{i \sigma \beta} + H.c.  \\
& &- \sum_{\alpha i \sigma} \mu_{\alpha}\;  n_{i \sigma \alpha}
+ \sum_{\alpha i} U_{\alpha \alpha}\; n_{i \uparrow \alpha} n_{i \downarrow \alpha}
\nonumber \\
& &+\sum_{\alpha \neq \beta, i} U_{\alpha \beta} \left[ n_{i \uparrow \alpha} n_{i \downarrow \beta}
- S^{+}_{i \alpha} S^{-}_{i \beta} + \Delta^{\dagger}_{i \alpha} \Delta_{i \beta}  \right]  ,
\nonumber
\end{eqnarray}
where the band index $\alpha,\beta=0,1$ for the [000] and [001] bands,
and the ``spin''\cite{fateofspin} index $\sigma = \uparrow, \downarrow$, and the site index  is $i$. 
$n_{i \sigma \alpha} = c^{\dagger}_{i \sigma \alpha} c_{i \sigma \alpha}$,
$S^+_{i \alpha} = c^{\dagger}_{i \uparrow \alpha} c_{i \downarrow \alpha}$, and
$\Delta_{i \beta} = c_{i \downarrow \beta} c_{i \uparrow \beta}$. The chemical potential
difference $\mu_0 - \mu_1 = \omega_0$ is the difference in energy between the two bands.
For simplicity, we here consider only a 1D system. This is achieved in the experiments of \cite{1D} by setting the laser amplitudes
$V_{0x} \ll V_{0y} = V_{0z}$,
thereby creating an uncoupled set of 1D tubes\cite{1D}, where excitations transverse to the tube axis are completely frozen. For higher
dimensions, because of the spatial anisotropy of the [001] orbital, the hopping
matrix element $t_{\alpha\beta}$ acquires spatial anisotropy also which complicates analysis.\cite{Girvin05,unpub}
 
This Hamiltonian has a similar form to those in solid state systems when multiple
orbitals are active in eg. transition metal oxides\cite{Tokura00}: the $- S^+ S^-$ term 
gives rise to
the Hund's rule in atomic physics, that favours spin alignment between different orbitals. The difference here in the optical lattice
is that  the atomic interactions are contact interactions, not the long range Coulomb
interactions for electrons in solids. Hence, the Hund's rule term
is of the same order of magnitude as the on-site repulsion term, unlike in solids. Also
unlike in solids, experimental preparation\cite{K04} dictates that the number 
of fermions for {\it each} spin species
is conserved separately.

In this paper, we study this model Eq.~\ref{H} at weak intersite tunneling and 
at zero temperature. We show  that when the Feshbach-tunable interactions are weak 
compared to the inter-band energy $\omega_0$, a "spin"-singlet\cite{fateofspin} of 
two fermions reside in each lattice well in the lowest Bloch band to form a band insulator. When the interactions are strong relative to $\omega_0$, the two fermions at each site reside in different bands with aligned "spin" forming an effective spin-1. This effective spin-1 alternates in sign between neighbouring sites to take advantage of second order virtual hopping of the fermions, thereby forming  a spin-1 Heisenberg  antiferromagnet, see Fig.\ref{fig:phase}. This is a correlated (Mott) insulator state as each band is only half-filled, and is reminiscent of the charge-transfer insulator in transition-metal compounds\cite{Zaanen85},
and excitonic insulators\cite{Littlewood04}.

The use of Feshbach resonance to populate higher Bloch bands of optical lattices promises to  open up much interesting orbital physics that may go beyond those models studied in the context of transition-metal oxides\cite{Tokura00}. 
For example, in 2D and 3D, there can be
orbital spatial ordering in a lattice with a single species of bosons\cite{Girvin05}. Duan \cite{Duan05} has also studied the effect of Feshbach resonance on fermions in optical lattices, but as that study was in a different regime, he found effective models with spin-1/2 antiferromagnetic correlations. 

{\it Model derivation}
The microscopic 1D Hamiltonian is:
\begin{eqnarray} \label{microH}
H = & & \sum_{\sigma} \int^L_0 dx \,  \frac{1}{2M} \left(\partial_x \psi^{\dag}_{\sigma}(x) \partial_x  \psi_{\sigma}(x) \right) \nonumber \\
& &  + \sum_{\sigma} \int^L_0 dx \, \left( V_{0x} \sin^2 k x - \mu \right) \: \psi^{\dagger}_{\sigma}(x)  \psi_{\sigma}(x) \nonumber\\
& &+ \frac{g}{2} \int^L_0 dx \: \psi^{\dag}_{\uparrow}(x) \psi^{\dag}_{\downarrow}(x) \psi_{\downarrow}(x) \psi_{\uparrow}(x) .
\end{eqnarray}
Since we are interested in the regime where at most two Bloch bands are occupied (roughly, $U< 2 \omega_0$), we expand the operator $\psi_{\sigma}(x)$ in Wannier functions:
$\psi_{\sigma}(x) = \sum_{\alpha=0,1} w_{i \alpha}(x) c_{i \sigma \alpha}$ ,
 where for a deep optical lattice, we approximate the Wannier functions by {\it local} harmonic oscillator  orbitals. I shall consider band 0 to be the ground state:
$w_{i 0}(x) =   \frac{1}{(\pi a_0^2)^{1/4}}  \exp \frac{(x-x_i)^2}{2 a^2_0}$, and with band 1 to be the first excited state:
$w_{i 1}(x) =   \frac{(-1)^{i}}{(\pi a_0^2)^{1/4}} \frac{\sqrt{2} (x-x_i)}{a_0} \exp \frac{(x-x_i)^2}{2 a^2_0}$. Note that these functions, unlike Wannier functions, are not orthogoal for different bands with respect to the hopping, hence there can be band-changing hopping  with $t_{01}=t_{10}$. In fact,
in general for band 1 to have a higher energy, $|t_{00}| < |t_{01}| < |t_{11}|$.  

 Then substituting the approximate Wannier functions
into Eq.~\ref{microH} gives the Hamiltonian of Eq.~\ref{H}, with the parameters:
$U_{00} = c_{00} U, \: U_{01} = c_{01} U, \; U_{11} = c_{11} U, \; 
U = \frac{a_{s} k}{\sqrt{\pi}} \left( \frac{4 \bar{V}}{E_r} \right)^{3/4}, \;
\bar{V} = \left( V_{0x} V_{0y} V_{0z} \right)^{1/3}$ . (We are giving the values for the actual 3D situation wherein the 1D tubes are embedded, {\it i.e.} $V_{0x} \ll V_{0y} = V_{0z}$.\cite{1D})
$a_{s}$ is the Feshbach resonance enhanced effective scattering length between the species, which can be tuned by magnetic field.\cite{inter-Fesh} In general $a_s$ will depend on the energy of the two scattering particles, and is renormalized
from the free space value by strong transverse confinement and the Feshbach 
resonance. In absence of calculations, we have left as parameters $c_{\alpha \beta}$
to denote this band dependence. As a crude guide, treating $a_s$ as 
energy-independent, $c_{00}=1, c_{01}=1/2, c_{11}=3/4$, and $a_s$ refers then to 
some average value, but with Feshbach and transverse confinement renormalization.
We only consider the situation where the broad Feshbach resonance is approached from 
the side where the scattering length is positive, with few if any molecules formed. Hence
sweeping the magnetic field towards  the Feshbach resonance effectively makes the
couplings $U_{\alpha \beta} \propto g = 4 \pi a_{s}/M >0 $ grows from $ U_{\alpha \beta} \ll \omega_0$ to $U_{\alpha \beta} \gtrsim \omega_0$.  We have approximated the interband energy
$\omega_0$ to be just the harmonic oscillator energy in the $x$ direction: $\omega_0 \approx \sqrt{ 4 V_{0x} E_r}$
with $E_r = k^2/2M $ the recoil energy for the optical lattice, $k = 2 \pi/\lambda$
where $\lambda$ is the laser wavelength, with the lattice spacing  $a=\lambda/2$.

\begin{figure}
 \begin{center}
 \includegraphics[width=\figwidth]{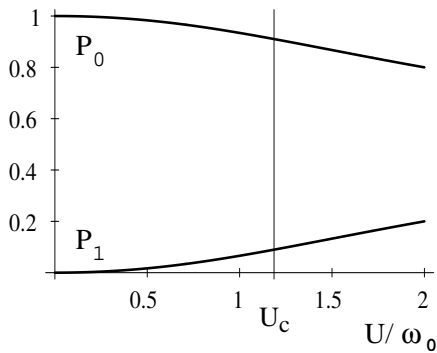}
 \caption{Probability of a singlet in band 0 ($P_0$: upper curve) or band 1 ($P_1$: lower curve)
 as a function of $U/\omega_0$, for the eigenstate $|s+\rangle$. The vertical line shows the location of $U_c$. Note that $|s+\rangle$ is the ground state  only for $U<U_c$, and for for $U\gtrsim 2 \omega_0$, higher bands start to be occupied.}\label{fig:occup}
 \end{center}
\end{figure}

\noindent{\it Strong coupling analysis}
K{\"o}hl {\it et.al.}'s experiment\cite{K04} is in the strong coupling limit, so we 
focus on $t_{\alpha \beta} \ll \omega_0 , U_{\alpha \beta}$. Thus, first diagonalise the Hamiltonian
Eq.~\ref{H} with $t_{\alpha \beta}=0$ to get the local spectrum with two fermions 
(one of each spin) per site:
\bea
E^{(t)}_{+}  &=&  -2 \mu_0 + \omega_0 , \quad \quad
 |t+\rangle = \frac{1}{\surd 2} 
\left( |\uparrow \; ; \downarrow \rangle + |\downarrow \; ; \uparrow \rangle \right) , \nonumber \\
E^{(t)}_{-}  &=&  -2 \mu_0 + \omega_0 + 2 U_{01} , \;\;
 |t-\rangle = \frac{1}{\surd 2} 
\left( |\uparrow \; ; \downarrow \rangle - |\downarrow \; ; \uparrow \rangle \right) , 
\nonumber \\
E^{(s)}_{\pm}  &=&  -2 \mu_0 + \omega_0 +\frac{U_{00}+U_{11}}{2} \nonumber\\
 & &  \;\; \mp 
\left[\left( \frac{U_{11}-U_{00}}{2} + \omega_0\right)^2 + U_{01}^2 \right]^{1/2}  \nonumber \\
|s+\rangle &=& a_1 |\uparrow\downarrow \; ; 0 \rangle + a_2 | 0 \; ; \uparrow\downarrow \rangle \nonumber\\
|s-\rangle &=& a_2 |\uparrow\downarrow \; ; 0 \rangle - a_1 | 0 \; ; \uparrow\downarrow \rangle
\eea
 The notation for the eigenstates is that the
left most slot is for band 0, separated by a $;$ from the band 1 slot. (We have dropped a
site index on the eigenstates.) $a_1, a_2$ are functions of $U_{\alpha \beta}$ and $\omega_0$ with $a_1^2 + a_2^2 =1$. In Fig.\ref{fig:occup}, we plot the probability
of a singlet in band 0 or band 1, using $c_{00}=1, c_{01}=1/2, c_{11}=3/4$.

Since $U_{\alpha \beta} >0$, the lowest energy is $E^{(s)}_{+}$ for 
$U_{\alpha\beta} < U^c_{\alpha\beta}$, or
$E^{(t)}_{+}$ when $U_{\alpha\beta} > U^c_{\alpha\beta}$. The two levels cross at 
$U_{\alpha \beta}^c(\omega_0)$ given by the implicit equation:
$\left(U^c_{01}\right)^2 = \left(U^c_{00}-\omega_0\right) \left(U^c_{11}+\omega_0\right)$ . Note that there is only one transition, since all the $U_{\alpha \beta} \propto U$. 
 Right at the transition, the parameters $a_1, a_2$
have the simple form:
$a_1 = - \left(\frac{U^c_{11}+ \omega_0}{U^c_{00}+U^c_{11}}\right)^{1/2}$, 
$a_2 =  \left(\frac{U^c_{00}- \omega_0}{U^c_{00}+U^c_{11}}\right)^{1/2}$ .
Thus, when $U_{\alpha\beta} < U^c_{\alpha\beta}$, at each site, the ground state
is $|s+\rangle$ with a singlet in band 0 mixed in with a bit of the 
singlet in band 1, while for $U_{\alpha\beta} > U^c_{\alpha\beta}$, 
the ground state
is a spin triplet with one fermion in each band: $|0\rangle \equiv |t+\rangle$. 
Note that  this triplet is degenerate with the other two members of the spin triplets:
$|U\rangle = |\uparrow \; ; \uparrow\rangle, \; |D\rangle = |\downarrow \; ; \downarrow\rangle$.
In K{\"o}hl {\it et al.}'s experiment, when the inter-species interaction is gradually turned up, the states 
$|U\rangle, |D\rangle$ cannot be reached at any sites if there are no hopping at all
between the sites. 

When the hopping is switched on, the effective low-energy Hamiltonian can be derived
using the usual strong coupling expansion to $O(t^2/U)$, where intermediate states have a site and its neighbour having three and one fermions respectively. We find\cite{unpub} that  for  $U_{\alpha \beta}<U_{\alpha \beta}^c$, the system is still a band insulator, 
while for $U_{\alpha \beta}>U_{\alpha \beta}^c$, the system is
now a Mott (correlated) insulator, with interesting spin dynamics which is that of
the spin-1 Heisenberg antiferromagnet:
\bea
H_{U>U_c} &=&  J \sum_i \left(\vec{S}_i \vec{S}_{i+1} -1 \right) , \\
 J &=& \frac{t_{00}^2}{U_{00}+U_{01}} + \frac{t^2_{11}}{U_{11}+U_{01}} \\
& & + t^2_{01} \left(\frac{1}{U_{00}+U_{01}-\omega_0}+\frac{1}{U_{11}+U_{01}+\omega_0} \right) \nonumber
\eea where $\vec{S}$ is a spin-1 operator. Spin-1 is involved because of the spin 
degeneracy of the triplet states. Unlike the 1D spin-1/2 antiferromagnet
which has no spin gap, the spin-1 case in 1D has a gap $\approx 0.41J$ (the Haldane gap\cite{Haldanegap,Haldanegap-num}).

At $U_{\alpha \beta} = U_{\alpha \beta}^c$, four states 
(the spin-1 triplet and $ |t+\rangle$) are now degenerate locally. The effective 
Hamiltonian at the transition point
correpsonds to a spin ladder, where each leg is a spin-1/2 Heisenberg 
antiferromagnet, with exchange couplings between the sites on a rung of the ladder
({\it i.e.} $\vec{S}^{(\alpha)}_i \vec{S}^{(\alpha)}_{i}$, $\alpha$ is now the ladder leg index), but also diagonally ({\it i.e.} $\vec{S}^{(1)}_i \vec{S}^{(2)}_{i+1}$, where 1,2 labels the two legs). Full details will be published elsewhere.\cite{unpub} We just point out here that
generically, such spin ladders have a spin gap\cite{LH00,G04}. Hence it is likely that
over the whole phase diagram as a function of $U$, the spin gap persists.

So far, we have focussed on the large $U$ limit and found only insulators. Clearly,
when $t_{\alpha \beta}\gg \omega_0, U_{\alpha \beta}$, the system will be a metal.
Thus in the whole phase diagram there should be metal-insulator transitions together
with some multicritical point. 
Work is in progress to explore this rich phase diagram.

\noindent{\it Experimental signatures} Both the band insulator and the Mott 
insulator have an energy gap to "charge" excitations, and interestingly, 
also for "spin" excitations. 
 However, for this band  to Mott insulator quantum phase transition,
there are spin gaps (and also charge gaps) all the way from 
$U_{\alpha\beta} < U^c_{\alpha\beta}$ to $U_{\alpha\beta} > U^c_{\alpha\beta}$ 
(except perhaps right at the transition?), so a spin gap (as might be measured
via 2-photon Raman transitions\cite{BZZ04,difft-v}) does not qualitatively 
distinguish the band vs. the Mott insulator. The smoking gun experiment 
would be to image the "up" spins populating
one sublattice while the "down" spins populate the other.

But even without direct spin and site dependent imaging, current experimental
probes such as   noise correlation\cite{Altman04,Folling05} from time-of-flight (TOF)
imaging may distinguish between the band insulator and the Mott insulator.
The 1D momentum distribution can be readily evaluated, assuming that deep in
the insulator phases, "charge" fluctuations are frozen. Thus, for the band insulator
phase where most weight is in the band 0, $n(Q_r) \sim \rho_{0} |w_0(Q_r)|^2 $,
while for the Mott insulator, with one fermion in each band at each site,
$n(Q_r) \sim \rho_{0} \left(|w_0(Q_r)|^2 + |w_1(Q_r)|^2\right)/2$. $Q_r = M r/t$ is the
continuum momentum with $r$ the imaging position relative to the initial distribution of cold atoms, and $t$ is the time of flight. This change in the filling of the bands
may have been observed in \cite{K04}.

To probe the "spin" spatial structure, one has to go to the noise correlation in the
TOF imaging, which measures 
$G(r,r') = \sum_{\sigma \sigma'} \langle n_{\sigma}(Q_r) n_{\sigma'}(Q_{r'}) \rangle- \langle n_{\sigma}(Q_r)\rangle \langle n_{\sigma'}(Q_{r'})\rangle$.\cite{Altman04}
A straightforward generalization to the two band case gives a number of terms that
show differences between the two types of insulators, but the key signature comes
from the static spin structure factor\cite{Altman04} part within $G(r,r')$. Because of the different spatial symmetry of the Wannier functions of the two bands, $G(r,r')$ does not directly probe the effective spin-1 correlations of the Mott insulator, but instead has information on individual band spin correlations:
\bea
G(r,r')
& \sim & -2\sum_{\alpha \beta} \sum_{n m} e^{i(Q_r-Q_{r'})(r_n-r_m)} \\
& &\times \; w_{\alpha}^*(Q_r) w_{\beta}(Q_r) w_{\beta}^*(Q_{r'}) w_{\alpha}(Q_{r'}) 
\nonumber\\
& &\times \left\langle \vec{S}_{\alpha n} \vec{S}_{\beta m} 
-\frac{1}{4} n_{\alpha n} n_{\beta m} \right\rangle  + \cdots \nonumber ,
\eea
where $\vec{S}_{\alpha n}$ is a spin-1/2 operator of band $\alpha$ at site $r_n=n a$.
(The $\cdots$ refer to other terms without lattice periodicity mentioned above, we shall henceforth ignore them.)
For the band insulator, since at each site there is a singlet in band 0,  there is no
spin-spin correlation contribution, and $G(r, r') \propto -\frac{1}{2} |w_0(Q_r)|^2 |w_0(Q_{r'})|^2
\sum_G \delta \left(Q_r - Q_{r'} + G\right)$, where $G$ is a reciprocal lattice vector
(integer multiples of $2 \pi/a$).
For the Mott insulator, assuming equal spin-spin correlation for the two bands and between bands, 
\bea
G(r, r') &\propto& -\left[ \frac{1}{2} \sum_G \delta \left(Q_r - Q_{r'} + G\right) \right. \nonumber \\
& & \left. + 2 \sum_G  f\left(Q_r - Q_{r'} + \frac{\pi}{a} + G\right) \right] \nonumber \\
& & \times \sum_{\alpha \beta} w_{\alpha}^*(Q_r) w_{\beta}(Q_r) w_{\beta}^*(Q_{r'}) w_{\alpha}(Q_{r'}) .
\eea Thus, just as for the spin-1/2 Heisenberg antiferromagnet, a sharp (only relatively so in 1D) peak of form $f(Q)$ (see eg. \cite{Altman04})  occurs in between the Bragg peaks at $G$, thanks to the doubling of the unit cell in real space for the antiferromagnet, see Fig.\ref{fig:phase}. The appearance
of this extra peak constitutes proof of the spin structure, while the existence of a spin gap
distinguishes between a spin-1 from a spin-1/2 antiferromagnet.

We have assumed so far that the system is homogeneous and large. In 
experiments\cite{K04}, the system consists of around $\sim 10^5$ fermions in a 
few thousand tubes, and the overall trapping potential leads to inhomogeneities in
occupation per site across the tube. This inhomogeneity can lead to phase 
coexistence\cite{R03} of a Mott insulator surrounded by a shell of superfluidity for the bosonic Hubbard model. 
Nevertheless, the bosonic Mott insulator has been observed\cite{G02,Folling05}, and 
we expect the same to be possible for the fermionic model here, as long as there is 
a large enough central region with commensurate filling of two fermions per site.
Also, in the experiment of \cite{K04}, the number of ``spin-up'' fermions equal the 
``down'' ones up to a few percent, and the spin-1 antiferromagnetic structure described above should persist in the Mott phase. 

We acknowledge inspiring conversations with T.L. Ho, M. Long, A. Nersesyan, M. Fabrizio, P. Julienne and P. Littlewood. A.F.H. is supported by EPSRC(UK).


\begin{thebibliography}{30}

\bibitem{G02}
M. Greiner \emph{et al.}, Nature (London) {\bf 415}, 39 (2002).
%
\bibitem{BCSBEC}  C.~A. Regal, M.~Greiner, and D.~S. Jin,  Phys.~Rev.~Lett. {\bf 92}, 040403, (2004), M. Zwierlein \emph{et al.}, \emph{ibid} {\bf 92}, 120403 (2004).
%
\bibitem{BCSBECtheory} E. Timmermans \emph{et al.}, Phys. Lett. A {\bf 258}, 228 (2001),
M. Holland \emph{et al.}, Phys.~Rev.~Lett. {\bf 87}, 120406 (2001). 
%
\bibitem{Jaksch98} D. Jaksch  \emph{et al.}, Phys.~Rev.~Lett. {\bf 81}, 3108 (1998).
%
\bibitem{K04}
M. K\"ohl  \emph{et al.},  Phys.~Rev.~Lett.~{\bf 94}, 080403 (2005).
%
\bibitem{Jason} R. B. Diener and T.-L. Ho, cond-mat/0507253 (2005).
%
\bibitem{Troyer05} H.G. Katzgraber, A. Esposito and M. Troyer, cond-mat/0510194 (2005).
%
\bibitem{fateofspin}
We use  ``spin'' to refer to internal (hyperfine or species) states of the fermions.
The real spin  of the  atoms is fixed by the experimental 
preparation of the  mixture.~\cite{K04,1D} "Charge" refers to both states.
%

\bibitem{1D}  
H. Moritz \emph{et al.}, Phys.~Rev.~Lett.  {\bf 91}, 250402 (2003), A.~F. Ho, M.~A. Cazalilla, and T. Giamarchi, \emph{ibid} {\bf 92}, 130405 (2004), 
H. Moritz \emph{et al.},   \emph{ibid}  {\bf 94},  210401 (2005).
%
\bibitem{Girvin05}  A. Isacsson and S. M. Girvin, cond-mat/0506622 (2005).
%
\bibitem{unpub} A. F. Ho, unpublished.
%
\bibitem{Zaanen85} J. Zaanen, G. A. Sawatzky and  J. W. Allen, Phys.~Rev.~Lett.~{\bf 55}, 418 (1985).
%
\bibitem{Littlewood04} P. B. Littlewood \emph{et al.}, J. Phys. Condens. Matter 
{\bf 16}, S3597 (2004)
%
\bibitem{Tokura00}  Y. Tokura and N. Nagaosa, Science {\bf 288}, 462 (2000). 
%
\bibitem{Duan05}  L. M. Duan, Phys.~Rev.~Lett.~{\bf 95}, 243202 (2005)
%
%
\bibitem{inter-Fesh}  C. A. Regal and D. S. Jin, Phys.~Rev.~Lett.~ {\bf 90}, 230404 (2003).
%
%
%
%
%
\bibitem{LH00}
E. Dagotto, J. Riera and D. Scalapino, Phys. Rev. B {\bf 45}, R 5744 (1992).
%
\bibitem{G04}
T. Giamarchi, \emph{Quantum Physics in One dimension}, Clarendon
Press (Oxford, UK, 2004); A.O. Gogolin, A.A. Nersesyan, A.M. Tsvelik,
\emph{ Bosonization and Strongly Correlated Systems}, Cambridge
University Press (Cambridge, UK, 1999); J. Voit, 
Rep.~Prog.~Phys.~{\bf 57}, 997 (1994). 
%
%
\bibitem{Haldanegap} F. D. M. Haldane,  Phys. Lett. {\bf 93 A}, 464 (1983).
%
\bibitem{Haldanegap-num} S. R. White and D. A. Huse, Phys. Rev. B {\bf 48}, 3844 
(1993), E. S. Sorensen and I. Affleck, Phys. Rev. Lett. {\bf 71}, 1633 (1993).  
%
\bibitem{BZZ04}
H.~P. Buchler, P. Zoller and W. Zwerger, Phys.~Rev.~Lett.~{\bf 93} 080401 (2004).
%
\bibitem{difft-v}
M.~A. Cazalilla, A.~F. Ho, and T. Giamarchi,  Phys. Rev. Lett. {\bf 95}, 226402 (2005).
%
\bibitem{Altman04} E. Altman, E. Demler and M.D. Lukin, Phys. Rev. A {\bf 70}, 013603
(2004), L. Mathey, E. Altman and A. Vishwanath, cond-mat/0507108 (2005).
%
\bibitem{Folling05} S. F\"olling {\it et al.}, Nature {\bf 434}, 481 (2005).
%
\bibitem{R03}
M. Rigol \emph{et al.}, Phys.~Rev.~Lett.~ {\bf 91}, 130403 (2003),
S. Wessel \emph{et al.}, Phys. Rev. A {\bf 70}, 053615 (2004).
%
\end{thebibliography}
\end{document}